\documentstyle[12pt, epsfig]{article}

\newcommand\be{\begin{equation}}
\newcommand\ee{\end{equation}}
\newcommand\bd{\begin{displaymath}}
\newcommand\ed{\end{displaymath}}
\newcommand\bea{\begin{eqnarray}}
\newcommand\eea{\end{eqnarray}}
\newcommand\nn{\nonumber}
\newcommand\lt{\left}
\newcommand\rt{\right}

\begin{document}

\begin{titlepage}
\title{\small\raggedleft{quant-ph/9903055}\\
\small\raggedleft{Fermilab Preprint: Pub-99/044-A}\\
\center\LARGE Higher Order Methods for Simulations \\
              on Quantum Computers}
\author{\small A. T. Sornborger \& E. D. Stewart,\\
        \small NASA/Fermilab Astrophysics Group,\\
        \small Fermi National Accelerator Laboratory,\\
        \small Box 500,\\
        \small Batavia, IL 60510-0500\\
        \small USA}
\date{\small March 15, 1999}
\end{titlepage}

\maketitle

\begin{abstract}

To efficiently implement many-qubit gates for use in quantum
simulations on quantum computers we develop and present methods
reexpressing $\exp[-i (H_1 + H_2 + \dots) \Delta t]$ as a product of
factors $\exp[-i H_1 \Delta t]$, $\exp[-i H_2 \Delta t]$, $\dots$
which is accurate to 3rd or 4th order in $\Delta t$. The methods we
derive are an extended form of symplectic method and can also be used
for the integration of classical Hamiltonians on classical
computers. We derive both integral and irrational methods, and find
the most efficient methods in both cases.

\end{abstract}
\thispagestyle{empty}

\newpage
\section{Introduction}

Quantum computers have generated much interest recently, largely due
to the result by Shor \cite{shor} that they can factor integers in
polynomial time. 

In a quantum computer the analog of a logical bit is the qubit.
The canonical example of a qubit is a quantum spin. A quantum spin
consists of two states, so a set of $n$ spins gives the quantum
computer a $2^n$-dimensional Hilbert space.

To perform a calculation, one initializes the qubits, and then applies
unitary logical gates to the qubits. Unitary logical gates are
realised in different fashions depending on the quantum computer
hardware, but they are all represented mathematically by a Hamiltonian
acting on the quantum state of the qubits. In a typical quantum
computer, technology restricts the Hamiltonian to act on a small
number of qubits at a time, maybe two or three. A calculation is then
built up of two- or three-qubit Hamiltonians, or gates, acting
sequentially on the qubits.

An important and difficult to realise requirement is that the qubits
maintain their coherence throughout an entire calculation. Maintaining
coherence in quantum computers is a problem which  has led to the
development of error correcting codes (see \cite{preskill} and
included references). These codes are possible due to the fact that
one does not need to know the state of a qubit in order to tell
whether an error has occurred. With some ingenuity, it is possible to
determine what kinds of errors have occurred during the course of a
calculation and to correct the errors as the calculation
proceeds. Simple error correction codes have already been shown to
work on small numbers of qubits \cite{zurek}.

Effort has also been put into developing algorithms which make use of
the quantum computer's power. Shor's algorithm showed that quantum
computers are more powerful than classical computers, since integers
cannot be factored in polynomial time on a classical computer, whereas
they can on a quantum computer. Grover has also devised a method for
searching a database in time proportional to the square root of the
number of items involved in the search \cite{grover}.

In addition to research into effective algorithms for use on quantum
computers, simulations of quantum systems have also been shown to be
possible in polynomial time \cite{simulate}. Indeed, this was the
first area for which it was proposed that quantum computers could
fundamentally be more powerful (i.e. much faster) than classical
computers \cite{feynman}.

This paper focuses on a problem which concerns simulational issues in
quantum computation. Essentially, we have developed methods for
reexpressing $\exp[-i (H_1 + H_2 + \dots) \Delta t]$ as a product of
factors $\exp[-i H_1 \Delta t]$, $\exp[-i H_2 \Delta t]$, $\dots$
which is accurate to 3rd or 4th order in $\Delta t$, as mentioned in
the abstract.

A simulation on a quantum computer consists of applying an operator
$\exp(-i H t)$ on a set of qubits, where $H$, the Hamiltonian of the
system of interest is suitably encoded (and discretized) to act on the
set of qubits. For many body systems, $H$ is a sum of terms. For
instance, in a one-dimensional Ising spin model, the Hamiltonian is
\begin{equation}
H = \sum_{n=1}^N \vec\sigma_n \cdot \vec\sigma_{n+1}
\end{equation}
where $N$ is the number of spins. Another example is the Hubbard
model Hamiltonian, used in the study of high-$T_c$ superconductivity,
which can be written \cite{abramslloyd} as the sum 
\be
H = \sum_{i=1}^m V_0 n_{i\uparrow} n_{i\downarrow} + \sum_{\langle i,
j \rangle \sigma} t_0 c_{i \sigma}^* c_{j \sigma} \label{ham}
\ee
where $V_0$ is the strength of the potential, and $n_{i\sigma}$ is the
operator for the number of fermions of spin $\sigma$ at site $i$. In
the second (kinetic energy) term, the sum $\langle i, j \rangle$
indicates all neighboring pairs of sites, $t_0$ is the strength of the
``hopping'', and $c_{i\sigma}$, $c_{i\sigma}^*$ are annihilation and
creation operators, respectively, of a fermion at site $i$ and spin
$\sigma$. 

These models give examples in which a large simulation on a classical
computer is impossible due to the exponential increase in the size of
the Hilbert space of the quantum system with the number of lattice
sites. A many-particle system can sometimes be simulated with fewer
qubits in first-quantized form \cite{abramslloyd}, but in either case
the Hamiltonian $H$ is a sum of terms, so our methods are equally
applicable to both cases.

If the quantum computer cannot act on all spins at once, as is the
case for quantum gate arrays \cite{qc}, it becomes necessary to find
ways of approximating the application of the above Hamiltonians with
few-qubit gates. To second order, for instance, we find that
\begin{equation}
\lt( e^{-i H_1 \Delta t} e^{-i H_2 \Delta t} \ldots e^{-i H_N \Delta t} \rt)
\lt( e^{-i H_N \Delta t} \ldots e^{-i H_2 \Delta t} e^{-i H_1 \Delta t} \rt)
= e^{-i 2(H_1 + H_2 + \ldots H_N) \Delta t
+ {\cal O} \lt[ \lt(\Delta t \rt)^3 \rt]}
\end{equation}
where $H_n$ are two-qubit gates (e.g. $\sigma_n \cdot \sigma_{n+1}$).

Below, we analyze the problem of deriving higher order methods of this
type, and find a set of equations which, once solved, give $3$rd and
$4$th order methods analogous to the above second-order method. We
solve and present the formulae for $3$rd and $4$th order methods as
well as developing methods for approximating expressions involving
commutators, $\exp([A, B])$. Since there is a large set of solutions
to our equations, we spend some effort trying to isolate and present
only the most efficient methods.

After presenting our methods, we then provide results from a simple
application to give the reader confidence that our methods are
correct. 

This kind of method has been investigated elsewhere, for different
reasons, in the context of Hamiltonian systems under the name
`symplectic' method. In the section on symplectic methods, we comment
on what we have done differently from other investigations of
symplectic methods, and why our methods are applicable to more general
problems. We then present a summary of our results in the conclusions
section.

We also provide appendices with useful expressions used in the
derivation of our results, and some proofs of statements in the text.

\newpage
\section{Mathematical Analysis and Equations}

We want to express $\exp\lt(\sum_{n=1}^N A_n\rt)$ as a product of individual
$\exp\lt(A_n\rt)$'s. In order to do this, we use the
Campbell-Baker-Hausdorff formula. The Campbell-Baker-Hausdorff formula
to 5th order is
\bd
\exp \lt( a A_1 \rt) \exp \lt( a A_2 \rt)
= \exp \lt[ a \lt( A_1 + A_2 \rt) + \frac{1}{2} a^2 A_{12}
+ \frac{1}{12} a^3 \lt( A_{112} + A_{221} \rt)
+ \frac{1}{24} a^4 A_{1221} \rt.
\ed
\be
\label{CH}
\lt. \mbox{} - \frac{1}{720} a^5
\lt( A_{11112} - 2 A_{21112} - 6 A_{11221}
 - 6 A_{22112} - 2 A_{12221} + A_{22221} \rt)
+ {\cal O} \lt( a^6 \rt) \rt]
\ee
where
\be
\label{A}
A_{kl \ldots mn} \equiv [ A_k , [ A_l , \ldots [ A_m , A_n ] \ldots ] ]
\ee

To find combinations of operators $\exp{A_i}$ which approximate
$\exp\lt(\sum_{n=1}^N A_n\rt)$ to some order it is first necessary to
choose a strategy for searching among the large number of possible
combinations. First of all, we cannot search brute force since there
are too many possible combinations, and, in any case, this would not
give us a formula valid for all $N$. Therefore, we pick a fundamental
ordering of the product of exponentials with parameters allowing for
transposes of the entire product as well as raising all the
exponentials in the fundamental unit to the same power.

By iterating the Campbell-Baker-Hausdorff formula, we can get an
expression for the fundamental unit in terms of a single exponential
\be\label{unit}
\lt( e^{a A_1} e^{a A_2} \ldots e^{a A_N} \rt)^\alpha
= \exp \sum_{p=1}^\infty \alpha a^p B_N^p
\ee
which defines the $B_N^p$ in terms of the $A_n$. Here, $p$ is an
exponent on $a$, and a label on the matrices $B_N^p$. We take
$\alpha = \pm 1$.

Now combine a succession $i = 1, \dots, I$ of fundamental units with
parameters $a_i$ and $\alpha_i$. Again iterating
Campbell-Baker-Hausdorff gives
\be\label{BX}
\exp \lt( \sum_{p=1}^\infty \alpha_1 a_1^p B_N^p \rt)
\ldots
\exp \lt( \sum_{p=1}^\infty \alpha_I a_I^p B_N^p \rt)
= \exp \lt( \sum_X \sigma_I^X B_N^X \rt)
\ee
The $B_N^X$ are generated from the $B_N^p$ by commutation. $X$
represents a label ${pq \ldots rs}$ where
\be
B_N^{pq \ldots rs} \equiv [B_N^p, [B_N^q, \ldots [B_N^r, B_N^s] \ldots
]]
\ee
$B_N^{pq \ldots rs}$ is of order $ p + q + \ldots + r + s $.
Up to 5th order we can take
\be
X \in \lt\{ 1; 2; 3, 12; 4, 13, 112; 5, 14, 23, 113, 221, 1112 \rt\}
\ee
These $B_N^X$ span the space of commutators of the $B_N^p$'s to 5th
order and for $N \geq 2$ they are independent. Formulae for the
$B_2^X$ in terms of $A_1$ and $A_2$ are given in Appendix A.2. The
$\sigma_I^X$ are defined in terms of $\alpha_i$ and $a_i$ by
Eq.~(\ref{BX}). Here again, the $X$'s are labels. 

\newpage
After some calculation, the Campbell-Baker-Hausdorff formula,
Eq.~(\ref{CH}), then gives
\be\label{sigp}
\sigma_I^p = \sum_{i=1}^I \alpha_i {a_i}^p
\ee
for $ p = 1, \ldots, 5 $,
\be\label{sigpq}
\sigma_I^{pq} = - \frac{1}{2} \sigma_I^p \sigma_I^q
+ \frac{1}{2} \sum_{i=1}^I {a_i}^{q-p}
\lt[ \lt( \sigma_i^p \rt)^2 - \lt( \sigma_{i-1}^p \rt)^2 \rt]
\ee
for $pq = 12, 13, 14, 23$,
\be\label{sigppq}
\sigma_I^{ppq} = - \frac{1}{2} \sigma_I^p \sigma_I^{pq}
- \frac{1}{6} \lt( \sigma_I^p \rt)^2 \sigma_I^q
+ \frac{1}{6} \sum_{i=1}^I {a_i}^{q-p}
\lt[ \lt( \sigma_i^p \rt)^3 - \lt( \sigma_{i-1}^p \rt)^3 \rt]
\ee
for $ppq = 112, 113, 221$ \footnote{For the purposes of calculating
$\sigma_I^{221}$, note that $\sigma_I^{21} \equiv -\sigma_I^{12}$.},
\be
\sigma_I^{1112} = - \frac{1}{2} \sigma_I^1 \sigma_I^{112}
- \frac{1}{3} \lt( \sigma_I^1 \rt)^2 \sigma_I^{12}
- \frac{1}{24} \lt( \sigma_I^1 \rt)^3 \sigma_I^2
+ \frac{1}{24} \sum_{i=1}^I a_i
\lt[ \lt( \sigma_i^1 \rt)^4 - \lt( \sigma_{i-1}^1 \rt)^4 \rt]
\ee

For approximations to $\exp\lt(\sum_{n=1}^N A_n\rt)$, we require all
$\sigma_I^X = 0$ except for $\sigma_I^1$ which is the coefficient of
$B_N^1 = \sum_{n=1}^N A_n$, and which should be greater than zero.

An interesting feature of 3rd order methods is that they require
inverses, \mbox{i.e.} they require backward time evolution during part
of the method.\footnote{After this work was completed, we
became aware that this point had also been noted in
\cite{otherwork}.} This can be proved using Eq.~(\ref{sigp}) with $p =
3$. It has no nontrivial solutions when the product $\alpha_i a_i$ is
positive for all $i$. Therefore for 3rd order methods, $\alpha_i
a_i$ must be negative for at least one $i$. From the left hand side of
Eq.~(\ref{unit}), we see that this means that there must be at least
one inverse. Similarly, from Eqs.~(\ref{sigp}) with $p = 3$ and $p =
4$ it can be proved that 4th order methods must have at least two
inverses.

In Appendix A.1, we also prove the fact that for integral solutions
$\sigma_I^1$ must be a multiple of 2 for a 2nd order method, a
multiple of 6 for a 3rd or 4th order method, and a multiple of 30 for
a 5th order method. Our searches suggest that the constraints on
$\sigma_I^1$ may actually be stronger; all 4th order methods that we
have found have $\sigma_I^1$ a multiple of 12, and we have not been
able to find any 5th order methods.

In Section~\ref{com}, we will consider approximation to
$\exp[A_1, A_2]$ for which we require all $\sigma_I^X = 0$ except
for $\sigma_I^2$.

\newpage
\section{Numerical Method for Solution of the Equations}   

We solve our equations for both integer and irrational solutions,
using different methods for each search.

Our method to solve Eqs.~(\ref{sigp}-\ref{sigppq}) for integers
is to pick values of $\alpha_i$ and $a_i$ and see if they satisfy the
equations. To do this we restrict the number of fundamental units by
fixing $I$. We also restrict the range of the $a_i$'s.

We start with Eq.~(\ref{sigp}), since, in this equation, order with
respect to $i$ does not matter. So, for a given set of values, we need
to consider only one permutation, not all permutations of the
values. This greatly reduces the size of the search.

Furthermore, we start by considering $p = 1$ and $3$, since it is only
the sign of $\alpha_i a_i$ that matters in these equations. This
means we can consider only the sign of the combination $\alpha_i
a_i$, and not the signs of $\alpha_i$ and $a_i$ individually. This
reduces the search further. These equations are particularly
restrictive for the case of few inverses.

After solving the $p = 1$ and $3$ equations, we introduce separate
signs for the $\alpha_i$'s and $a_i$'s and solve the equation with $p
= 2$, and $p = 4$ for the 4th order case.

Finally, into the restricted set of solutions to Eq.~(\ref{sigp}) we
introduce permutations of the $\alpha_i$'s and $a_i$'s with respect to
the index $i$ and solve Eqs.~(\ref{sigp}-\ref{sigppq}).

We solved Eqs~(\ref{sigp}) and~(\ref{sigpq}) analytically to find the
unique shortest irrational 3rd order method. To find 4th order
irrational methods, we made a symmetric ansatz and solved
Eqs.~(\ref{sigp}-\ref{sigppq}) analytically to find the shortest
symmetric irrational 4th order methods. We checked numerically, using
the globally convergent technique prescribed in \cite{numrec}, that
these are all the shortest irrational 4th order methods.

The methods are presented in Section~(\ref{formulae}).

\section{Criteria for Selecting Among the Solutions \label{criteria}}

With our strategy for finding solutions to
Eqs.~(\ref{sigp}-\ref{sigppq}) we find a larger number of solutions
than we can easily present. We need to select solutions to present and
we also want to present solutions which are in some sense optimal. To
do this, we consider the form of the operator resulting from a given
method
\be
\prod_{j=1}^I \lt( e^{-i a_j A_1 \,\Delta t} \, e^{-i a_j A_2 \, \Delta t}
\ldots e^{-i a_j A_N \, \Delta t} \rt)^{\alpha_j}
= \exp \lt[ -i \sigma_I^1 \sum_{n=1}^N A_n \,\Delta t
 + r (-i \,\Delta t)^{o+1} \rt]
\ee
where $\Delta t \ll 1$ is a time step, $o$ is the order of the method,
and
\be\label{r}
r = \sum_X \sigma_I^X B_N^X
\ee
where $X \in \{4, 13, 112\}$ for a 3rd
order method and $X \in \{5, 14, 23, 113, 221, 1112\}$ for a 4th order
method.

$r$ is an error which takes values in the vector space of the
commutators for which we do not have a metric. Therefore, we make the
{\it ad hoc\/} choice of basis that is given in Appendix A.3. This
allows us to replace $r$ by a single real scalar $R$ as is also
described in Appendix A.3. The error from the method can then be taken
to be
\be\label{E}
E = n R \, \Delta t^{o+1}
\ee
where $n$ is the number of times we apply the approximate method.

If the physical time we want to simulate is $T_p$, then
\be\label{TP}
T_p = n D \Delta t
\ee
where $D \equiv \sigma_I^1$ is given by the method.

The computer time it takes for a given simulation can be written
\be 
T_c = n I N t_g + n L N t_s
\ee
where $I$ is the number of fundamental units in the method and $N$ is
the number of terms in a unit, $t_g$ is the time it takes to make the
gate change,
\be
L \equiv \sum_{i = 1}^I |a_i|
\ee
so that $L N$ is the total time the gates are applied for in the
method, and $t_s$ is the time each individual gate is applied for. The
time an individual gate is applied for will be $t_s = b \, \Delta t$,
where $b$ is a proportionality constant dictated by the actual
couplings in the quantum computer hardware.

Using Eqs.~(\ref{E}) and~(\ref{TP}), the computer time can be
rewritten 
\be
  T_c = \lt\{ \lt( \frac{T_p^{o+1}}{E} \rt)^{\frac{1}{o}}
        \lt( \frac{I}{D} \rt) \lt( \frac{R}{D}
        \rt)^{\frac{1}{o}} t_g + \frac{LbT_p}{D} \rt\} N
\ee

There are two possible limits to this equation. If $\Delta t$ can be
made very small (from the hardware point of view), then making the
error small forces the computer time to be dominated by gate
switching. In this case, we want the factor
\be
Z = (I/D) (R/D)^{1/o}
\ee
to be small. 

If there is a lower limit to $\Delta t = \epsilon$, and it is reached
before the computer time is gate switching dominated, then the
computer time may be dominated by gate application. In this second
limit, we want $L/D$ small, and to minimize the error $E$, we want
$(R/D)(\Delta t)^o$ small. In this limit, each gate can only be
applied for an integral number of the minimum timestep
$\epsilon$. Thus, to use an irrational method, one must approximate
the method by an integral method containing large integers, and so
with a large $D$. Because $\Delta t = \epsilon$ is fixed, the error
$E$ goes like $R/D \propto D^o$, and thus is large for irrational
methods. We thus do not consider irrational methods in this limit.

To summarize, we want methods with small $L/D$ and $R/D$.

\section{3rd and 4th Order Formulae for $\exp \lt( \sum_{n=1}^N A_n
\rt)$ \label{formulae}}

{}From this analysis, we want to choose methods for which
$Z$, or $L/D$ and $R/D$ are small.
Below we list the methods and their properties. We use the notation
\be
  (\alpha a)
\ee
to represent
\be
  \lt( e^{a A_1} e^{a A_2} \ldots e^{a A_N} \rt)^\alpha
\ee
if $\alpha = 1$, and 
\be
  (\alpha a)^T
\ee
to represent
\be
  \lt( e^{a A_1} e^{a A_2} \ldots e^{a A_N} \rt)^\alpha
\ee
if $\alpha = -1$. So, for example, the 2nd order method
\be
  \lt( e^{A_1} e^{A_2} \ldots e^{A_N} \rt) \lt( e^{A_N} \ldots e^{A_2}
    e^{A_1} \rt) = \lt( e^{A_1} e^{A_2} \ldots e^{A_N} \rt) \lt(
    e^{-A_1} e^{-A_2} \ldots e^{-A_N} \rt)^{-1}
\ee
is represented by
\be
  (1)(1)^T
\ee

Note that the transpose of any method gives another equivalent method,
as does permuting the entries in the fundamental unit.

For odd order methods, the residue has an odd number of brackets in
the commutators. So, because the transpose of an individual bracket is
minus that bracket,
\be
\lt( \mbox{odd order method} \rt)
\lt( \mbox{same odd order method transpose} \rt)
\ee
gives a method of one order higher. For example, we can make a 4th
order method from a 3rd order method, or a 6th order method from a 5th
order method.

\subsection{Integer solutions}

The 3rd order integer methods that we have selected using the criteria
of Section~\ref{criteria} are given below.

\vspace{2ex}

\begin{tabular}{|c|l|}
\hline
& 3rd Order Methods \\
\hline
${\cal Z}_3^1$ & $
(1)^T(1)(1)(1)(1)^T(-2)^T(1)(1)(1)
$ \\
${\cal Z}_3^2$ & $
(1)^T(4)(2)(-5)^T(2)^T(3)(2)(2)^T(1)
$ \\
${\cal Z}_3^3$ & $
(1)^T(2)(2)(-3)^T(1)^T(2)(1)^T
$ \\
${\cal Z}_3^4$ & $
(3)(-4)^T(1)(3)(2)^T(1)
$ \\
${\cal Z}_3^5$ & $
(5)^T(7)(12)(-13)^T(1)
$ \\
\hline
\end{tabular}\label{3rdtab}

\vspace{2ex}

\begin{tabular}{|c|rrrrrr|}
\hline
& $D$ & $L$ & $I$ & $L/D$ & $R/D$ & $Z$ \\
\hline
${\cal Z}_3^1$ & 6 & 10 & 9 & 1.67 & 0.2 & 0.9 \\
${\cal Z}_3^2$ & 12 & 22 & 9 & 1.83 & 0.6 & 0.6 \\
${\cal Z}_3^3$ & 6 & 12 & 7 & 2.00 & 0.4 & 0.9 \\
${\cal Z}_3^4$ & 6 & 14 & 6 & 2.33 & 1.7 & 1.2 \\
${\cal Z}_3^5$ & 12 & 38 & 5 & 3.17 & 98.8 & 1.9 \\
\hline
\end{tabular}

\vspace{2ex}

\begin{flushleft}
And the 4th order integer methods are
\end{flushleft}

\vspace{2ex}

\begin{tabular}{|c|l|}
\hline
& 4th Order Methods \\
\hline
${\cal Z}_4^1$ & $
(1)^T(1)(1)^T(-2)(1)^T(1)^T(1)^T(1)^T(1)(1)^T(1)(1)(1)(1)(-2)^T(1)(1)^T(1)
$ \\
${\cal Z}_4^2$ & $
(1)^T(2)(1)^T(-3)^T(2)(2)(1)(2)^T(2)^T(-3)(2)^T(1)(1)(1)^T
$ \\
${\cal Z}_4^3$ & $
(1)^T(2)(3)^T(1)^T(-4)(3)^T(3)(-4)^T(1)(3)(2)^T(1)
$ \\
${\cal Z}_4^4$ & $
(6)^T(-7)(1)^T(1)(5)^T(5)(1)^T(1)(-7)^T(6)
$ \\
\hline
\end{tabular}\label{4thtab}

\vspace{2ex}

\begin{tabular}{|c|rrrrrr|l|}
\hline
& $D$ & $L$ & $I$ & $L/D$ & $R/D$ & $Z$ \\
\hline
${\cal Z}_4^1$ & 12 & 20 & 18 & 1.67 & 0.6 & 1.3 \\
${\cal Z}_4^2$ & 12 & 24 & 14 & 2.00 & 0.8 & 1.1 \\
${\cal Z}_4^3$ & 12 & 28 & 12 & 2.33 & 4.6 & 1.5 \\
${\cal Z}_4^4$ & 12 & 40 & 10 & 3.33 & 50.2 & 2.2 \\
\hline
\end{tabular}

\subsection{Irrational solutions}

The equations that we have derived can be solved for irrational
solutions. We have been able to find the shortest 3rd order method
analytically. It can be proven to be unique. It is
\be
  \lt( a_1 \rt) \lt( -a_2 \rt)^T \lt( -a_3 \rt)^T \lt( a_4 \rt)
\ee
where
\bea
  a_1 & = & 1 \nn \\
  a_2 & = & -\frac{1}{6} \lt( 5 - \sqrt{13} \, + 2 \sqrt{5 +
           2\sqrt{13}} \rt) \nn \\
  a_3 & = & 1/\lt( 1 + a_2 \rt) \nn \\
  a_4 & = & -a_2 \lt( 1 + a_2 \rt)/ \lt( 3 + 2 a_2 \rt)
\eea
Renormalising to give $\sigma_I^1 = 1$, we have method ${\cal R}_3^1$
\be
\begin{array}{rcr}
  a_1 & = &  0.451525513208585723409578820 \\
  a_2 & = & -0.630880954030002500791663663 \\
  a_3 & = & -1.136710925213995714728206549 \\
  a_4 & = & -1.219117392452583938929449032 
\end{array}
\ee
accurate to 27 decimal places. This method has $Z = 1.7$.

From this 3rd order method, we can generate the 4th order method
\be
  \lt( a_1 \rt) \lt( -a_2 \rt)^T \lt( -a_3 \rt)^T \lt( a_4 \rt)
  \lt( a_4 \rt)^T \lt( -a_3 \rt) \lt( -a_2 \rt) \lt( a_1 \rt)^T
\ee

We have also found short fourth order methods. We assume a solution of
symmetric form, using the ansatz $\alpha_{I-i+1} = -\alpha_i$ and
$a_{I-i+1} = -a_i$. For $I = 6$, this leaves us with the equations
\be
  \sum_{i=1}^3 \alpha_i a_i = \frac{1}{2}
\ee
\be
  \sum_{i=1}^3 \alpha_i a_i^3 = 0
\ee
\be
  \sum_{i=1}^3 \lt[ a_i^3 + 2\alpha_i a_i^2 \lt( \sigma_3^1 -
    \sigma_i^1 \rt) \rt] = 0
\ee
to solve.

Combining equations and setting $\alpha_1 = 1$, we find solutions of
the form
\bea
  a_1 &=& \frac{1}{2 \lt( \alpha_2 x + \alpha_3 y + 1 \rt)} \nn \\
  a_2 &=& x a_1 \nn \\
  a_3 &=& y a_1
\eea
where
\be
  y = -\alpha_3 \lt( \alpha_2 x^3 + 1 \rt)^{1/3}
\ee
and $x$ has four possible values depending on the
$\alpha_i$'s. From our ansatz, $\alpha_4 = -\alpha_3$, $\alpha_5 =
-\alpha_2$, $\alpha_6 = -\alpha_1$, $a_4 = -a_3$, $a_5 = -a_2$ and
$a_6 = -a_1$.

For $\alpha_2 = -\alpha_3 = -1$, $x = -1$ giving the method ${\cal R}_4^1$
\be
\begin{array}{rcrcr}
  a_1 & = &  \frac{1}{4} \lt(2 + \sqrt{2} \rt) 
    & \simeq &  0.675603595979828817023843904 \\
  a_2 & = & -\frac{1}{4} \lt(2 + \sqrt{2} \rt) 
    & \simeq & -0.675603595979828817023843904 \\
  a_3 & = & -\frac{1}{2} \lt(1 + \sqrt{2} \rt) 
    & \simeq & -0.851207191959657634047687809 
\end{array}
\ee
This method has been found previously by Yoshida \cite{yoshida} in the
two-operator case, we see here that it is also a method for an
arbitrary sum of non-commuting operators. This method has $Z = 2.67$.

For $\alpha_2 = \alpha_3 = -1$, $x$ is the solution of
\be
  x^5 + 3x^4 + 3x^3 - 3x -3 = 0
\ee
giving the method ${\cal R}_4^2$
\be
\begin{array}{rcr}
  a_1 &=& -1.075035037431900314780251056 \\
  a_2 &=& -1.024607977441460486144230714 \\
  a_3 &=& -0.550427059990439828636020342 
\end{array}
\ee
This method has $Z = 2.53$, and is slightly better than the above
method of Yoshida.

For $\alpha_2 = -\alpha_3 = 1$, $x$ is the solution of
\be
  2x^5 + 3x^3 + 3x^2 + 3 = 0
\ee
giving the method ${\cal R}_4^3$
\be
\begin{array}{rcr}
  a_1 &=&  0.938925888779098070854126976 \\
  a_2 &=& -1.002122279211397565598116356 \\
  a_3 &=& -0.563196390432299494743989380
\end{array}
\ee
This method has $Z = 3.56$.

And finally, for $\alpha_2 = \alpha_3 = 1$, $x$ is the solution of
\be
  x^9 + 3x^7 + x^6 + 3x^5 + 3x^4 + 3x^2 +1 = 0
\ee
giving the method ${\cal R}_4^4$
\be
\begin{array}{rcr}
  a_1 &=&  1.087752928204421689142747144 \\
  a_2 &=& -1.131212302433601022822197398 \\
  a_3 &=&  0.543459374229179333679450254
\end{array}
\ee
This method has $Z = 4.39$.

We also searched numerically for other irrational solutions and found
no short asymmetric solutions (\mbox{i.e.} shorter than the symmetric
solutions found analytically).

\section{An Efficient Technique for Deriving Sub-optimal
\\ Higher Order Methods}

The technique for finding higher order methods described above used a
first order method as a fundamental unit. We can also use higher order
methods as fundamental units. This makes it easier to derive very high
order methods, but the methods will be sub-optimal in the sense that
we only generate a restricted set of solutions, which is unlikely to
contain the method that is optimal with respect to any given criteria.

The technique of using higher order fundamental units works as
follows:

The method of order $o$ from which we form the fundamental unit is
\be
  \prod_{i = 1}^I \lt( e^{a_i A_1} \ldots e^{a_i A_N} \rt)^{\alpha_i}
\ee
and the fundamental unit is
\be
  \lt[ \prod_{i=1}^I \lt(e^{a_i b A_1} \ldots e^{a_i
    b A_N} \rt)^{\alpha_i} \rt]^\beta = \exp\lt( \beta b \sigma_I^1
    \sum_{n=1}^N A_N + \beta b^{o+1} r \rt)
\ee

Combining a succession $j = 1, \ldots, J$ of fundamental units with
parameters $b_j$ and $\beta_j$ gives
\be
  \prod_{j=1}^J \lt[ \prod_{i=1}^I \lt(e^{a_i b_j A_1} \ldots e^{a_i
    b_j A_N} \rt)^{\alpha_i} \rt]^{\beta_j} =  \exp\lt( \sum_{j=1}^J\beta_j
    b_j  \sigma_I^1 \sum_{n=1}^N A_N + \sum_{j=1}^J \beta_j
    b_j^{o+1} r \rt)
\ee
Therefore, to obtain a method of order $o + 1$, we require
\be
  \sum_{j=1}^J\beta_j b_j > 0
\ee
and
\be
  \sum_{j=1}^J \beta_j b_j^{o+1} = 0
\ee

This technique can be iterated to get arbitrarily high order methods.

As an example, we start with the first order method
\be
  \lt( 1 \rt)
\ee
by transposing, we get the second order method
\be
  \lt( 1 \rt) \lt( 1 \rt)^T
\ee
now, we solve the equations
\be\label{firstorder}
  \sum_{j=1}^J\beta_j b_j > 0
\ee
and
\be\label{thirdorder}
  \sum_{j=1}^J \beta_j b_j^3 = 0
\ee

A simple solution to Eqs.~(\ref{firstorder} and \ref{thirdorder}) is
$2^3 = 1^3 \times 8$. Ordering is not dictated by the solution, so we
choose the method which is its own transpose and hence 4th order
accurate
\be
  \lt[ \lt( 1 \rt) \lt( 1 \rt)^T \rt]^4 \lt[ \lt( -2 \rt) \lt( -2
    \rt)^T \rt] \lt[ \lt( 1 \rt) \lt( 1 \rt)^T \rt]^4
\ee

Again, we solve the equations
\be
  \sum_{k=1}^K\gamma_k c_k > 0
\ee
and
\be\label{fifthorder}
  \sum_{k=1}^K \gamma_k c_k^5 = 0
\ee
which have the simple solution $2^5 = 1^5 \times 32$. Again, choosing
the ordering so that the method is its own transpose, gives the 6th
order method
\bea
  \lt\{ \lt[ \lt( 1 \rt) \lt( 1 \rt)^T \rt]^4 
       \lt[ \lt( -2 \rt) \lt( -2 \rt)^T \rt] 
       \lt[ \lt( 1 \rt) \lt( 1 \rt)^T \rt]^4 \rt\}^{16}
  \lt[ \lt( -2 \rt) \lt( -2 \rt)^T \rt]^4 
  \lt[ \lt( 4 \rt) \lt( 4 \rt)^T \rt]
  \lt[ \lt( -2 \rt) \lt( -2 \rt)^T \rt]^4 \nn \\
  \lt\{ \lt[ \lt( 1 \rt) \lt( 1 \rt)^T \rt]^4 
       \lt[ \lt( -2 \rt) \lt( -2 \rt)^T \rt] 
       \lt[ \lt( 1 \rt) \lt( 1 \rt)^T \rt]^4 \rt\}^{16}
\eea
where $I = 594$.

\section{4th and 5th Order Formulae for $\exp([A_1, A_2])$}
\label{com}

As a byproduct of our analysis, we can also use
Eqs.~(\ref{sigp}-\ref{sigppq}) to search for approximations to gates
involving commutators. To do this, we set $\sigma_I^2 > 0$ and
$\sigma_I^X = 0$ for $X \neq 2$. An approximation for a gate involving
a commutator may be useful if only a subset of the generators of a
particular group is available in hardware, but a given algorithm needs
another generator of the group. For instance, if $\exp{(-i \sigma_x
\Delta t)}$ and $\exp{(-i \sigma_y \Delta t)}$ are available in
hardware, but $\exp{(-i \sigma_z \Delta t)}$ is not, then we need a
way to generate $\exp{\lt(-\frac{1}{2} \lt[\sigma_x,
\sigma_y\rt]\Delta t\rt)}$.

After some searching, we have been able to find one method for
$\exp{[A,B]}$ to fourth order. It is
\bea
  (-2)^T(2)^T \lt[(-1)(1)\rt]^{12} \lt[(1)(-1)\rt]^4
\eea
with residuals

\begin{flushleft}
\begin{tabular}{|r|r|r|r|r|r|r|}
\hline
$\rho_{12}$ 
& $\rho_{11112}$ & $\rho_{21112}$ & $\rho_{11221}$ 
& $\rho_{22112}$ & $\rho_{12221}$ & $\rho_{22221}$ \\
\hline
12.0 & 1.0 &  2.0 &  0.0 & 0.0 & -2.0 & -1.0 \\
\hline
\end{tabular}
\end{flushleft}

This method can be combined with its transpose to give a 5th order
method.

\section{A Simple Application}

To illustrate our methods, we have applied first, second, third and
fourth order methods to the exactly soluble operator
\be
  e^{-i \Delta t (\sigma_x + \sigma_y + \sigma_z)} = 
  \lt( \begin{array}{cc}
    \cos\lt( \sqrt{3} \, \Delta t \rt) 
        - i \frac{1}{\sqrt{3}}\sin\lt( \sqrt{3} \, \Delta t \rt)
    & -(i + 1) \frac{1}{\sqrt{3}}\sin\lt( \sqrt{3} \, \Delta t \rt)\\
    -(i - 1) \frac{1}{\sqrt{3}}\sin\lt( \sqrt{3} \, \Delta t \rt)
    &  \cos\lt( \sqrt{3} \, \Delta t \rt) 
        + i \frac{1}{\sqrt{3}}\sin\lt( \sqrt{3} \Delta t \rt)
  \end{array} \rt)
\ee

\begin{figure}[ht]
\centerline{\psfig{figure=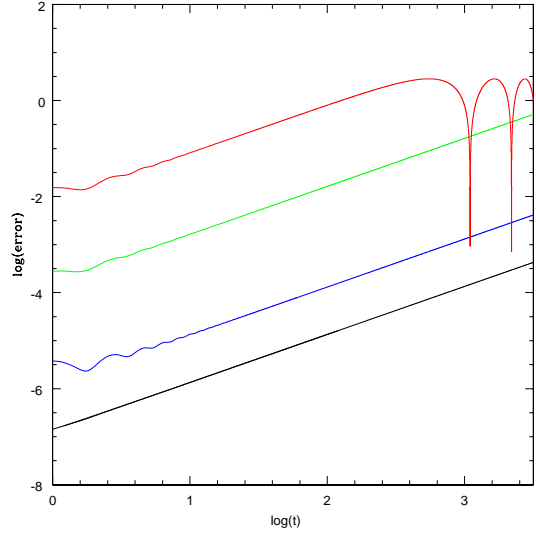,height=3.0in,width=3.0in}}
\caption{Here, we plot log(error) vs. log(time). Error is calculated
according to Eq.~(\ref{error}). The lines from top to bottom
correspond to the 1st, 2nd, 3rd and 4th order methods of
Eqs.~(\ref{firstordermethod}), (\ref{secondordermethod}),
(\ref{thirdordermethod}) and
(\ref{fourthordermethod}). \label{singlespin}}
\end{figure}

We used the first order method
\be
  {\cal Z}_1^1 = (1) =
\lt(
e^{-i \Delta t \sigma_x} e^{-i \Delta t \sigma_y} e^{-i \Delta t \sigma_z}
\rt) \label{firstordermethod}
\ee
the second order method
\be
  {\cal Z}_2^1 = (1)(1)^T = 
\lt(
e^{-i \Delta t \sigma_x} e^{-i \Delta t \sigma_y} e^{-i \Delta t \sigma_z}
\rt)
\lt(
e^{-i \Delta t \sigma_z} e^{-i \Delta t \sigma_y} e^{-i \Delta t \sigma_x}
\rt) \label{secondordermethod}
\ee
the third order method
\bea
{\cal Z}_3^1 &=& \lefteqn{(1)^T(1)(1)(1)(1)^T(-2)^T(1)(1)(1)} \nn \\
&=&
\begin{array}{l}
\lt(
e^{-i \Delta t \sigma_z} e^{-i \Delta t \sigma_y} e^{-i \Delta t \sigma_x}
\rt)
\lt(
e^{-i \Delta t \sigma_x} e^{-i \Delta t \sigma_y} e^{-i \Delta t \sigma_z}
\rt)
\lt(
e^{-i \Delta t \sigma_x} e^{-i \Delta t \sigma_y} e^{-i \Delta t \sigma_z}
\rt) \\
\lt(
e^{-i \Delta t \sigma_x} e^{-i \Delta t \sigma_y} e^{-i \Delta t \sigma_z}
\rt)
\lt(
e^{-i \Delta t \sigma_z} e^{-i \Delta t \sigma_y} e^{-i \Delta t \sigma_x}
\rt)
\lt(
e^{2 i \Delta t \sigma_z} e^{2 i \Delta t \sigma_y} e^{2 i \Delta t \sigma_x}
\rt) \\
\lt(
e^{-i \Delta t \sigma_x} e^{-i \Delta t \sigma_y} e^{-i \Delta t \sigma_z}
\rt)
\lt(
e^{-i \Delta t \sigma_x} e^{-i \Delta t \sigma_y} e^{-i \Delta t \sigma_z}
\rt)
\lt(
e^{-i \Delta t \sigma_x} e^{-i \Delta t \sigma_y} e^{-i \Delta t \sigma_z}
\rt) 
\end{array}
\label{thirdordermethod}
\eea
and similarly for the fourth order method
\be
  {\cal Z}_4^1 = 
  (1)^T(1)(1)^T(-2)(1)^T(1)^T(1)^T(1)^T(1)(1)^T(1)(1)(1)(1)(-2)^T(1)(1)^T(1)
  \label{fourthordermethod}
\ee

As a measure of the error, we take the difference between the
$\sigma_x$, $\sigma_y$ and $\sigma_z$ components of the exact solution
and our methods, $\Delta\sigma_x$, $\Delta\sigma_y$ and
$\Delta\sigma_z$. We then calculate the error
\be
  E = \sqrt{\lt( \Delta\sigma_x \rt)^2 + \lt( \Delta\sigma_y \rt)^2 
               + \lt( \Delta\sigma_z \rt)^2} \label{error}
\ee

In Fig.~(\ref{singlespin}), we plot the logarithm of the error as a
function of the logarithm of the time that the system was evolved
for. The first order method results are uppermost and higher order
results lie underneath each other with fourth order results being the
lowermost plotted. $\Delta t = 0.01$ for all methods.

Notice that the first order error oscillates once it reaches order
$1$. The rest of the errors remain small throughout the simulation,
with the fourth order error remaining below $10^{-3}$ for the entire
evolution.

The error for all methods goes as $n R \lt( \Delta t \rt)^{o+1}$,
where $n$ is the number of times the method has been
applied. Therefore, $\log E = \log \lt[ R \lt( \Delta t \rt)^{o + 1}
\rt] + \log n$. For $\Delta t = 0.01$, this makes the $y$-intercept
decrease by order $-2$ as the order of the method increases. Since the
time evolved is proportional to $n$, the slope of the errors is $1$
for all methods.

\section{Symplectic Methods}

In the study of classical Hamiltonian systems, we can cast the
evolution of the coordinates $q_i$ and momenta $p_i$ of fields or
particles in the same language as we have done above for quantum
systems.

Write $z = (q_i, p_i)$. Then the Hamilton equations for
the system are
\be
  \dot z = \{z, H\}
\ee
where $\{a, b\}$ is a Poisson bracket. Now, define $D_H z \equiv \{z,
H\}$. The Hamilton equations become
\be
  \dot z = D_H z
\ee
The formal solution to these equations is then,
\be
  z(t) = e^{D_H t} z_0
\ee
Often, $D_H$ can be separated into kinetic and potential parts
$D_H = D_K + D_V$. In this case, we have the formal solution
\be
 z(t) = e^{(D_K + D_V) t} z_0 \label{symplec}
\ee
Typically, symplectic methods approximate the above case
(\ref{symplec}), in which there are only two operators in the
exponential. Symplectic methods for two operators exist up to 8th
order in the expansion \cite{yoshida}.

In our work, we have developed methods to approximate the case where
there are an arbitrary number of operators in the exponential. This is
important for simulations on both quantum and classical computers,
since there can often be more than two terms which do not commute in
the Hamiltonian.

For example, any Hamiltonian of the form
\bea
  H = g_{ij}(q) p_i p_j + V(q)
\eea
where $g_{ij}$ and $V$ are functions of the $q_i$'s, can have an
arbitrary number of terms which do not commute with each other.

A simple example of a quantum system where extra terms in the sum are
necessary is an Ising spin system with next-nearest neighbor
interactions. Here, the Hamiltonian becomes
\be
  H = \sum_{i=1}^N \lt( \sigma_i \cdot \sigma_{i+1} 
             + \sigma_i \cdot \sigma_{i+2} \rt)
\ee
In this Hamiltonian, none of the terms 
$\sigma_{i-2} \cdot \sigma_{i}$, 
$\sigma_{i-1} \cdot \sigma_{i}$, 
$\sigma_{i} \cdot \sigma_{i+1}$ or
$\sigma_{i} \cdot \sigma_{i+2}$ commute. Therefore, for this system,
we can arrange the Hamiltonian to have, at best, four terms which do
not commute with each other.

\section{Conclusions}

The object of this paper has been to provide higher order
approximation methods for operators of the form $\exp{\sum_{i=1}^N
A_N}$ in terms of operators of the form $\exp{(A_1)}$, $\exp{(A_2)}$,
$\ldots$, $\exp{(A_N)}$. We have focused on approximation methods of
this kind since they are particularly useful in quantum many-particle
simulations for which the discretised Hamiltonian on a quantum
computer takes the form of an exponential of a sum of non-commuting
terms.

To find higher order methods, we have derived and solved equations for
methods up to 4th order. We find that the equations give a large
number of methods, so we have selected a small number of them based
on what seem to us to be reasonable criteria and presented them
above.

As a by-product of our search, we have also been able to find higher
order approximation methods for operators of the form $\exp{[A,B]}$ in
terms of operators of the form $\exp{(A)}$ and $\exp{(B)}$. These may
be useful for quantum gates where $\exp{(A)}$ and $\exp{(B)}$ are
available in hardware, but the gate $\exp{[A,B]}$ is desired for some
particular algorithm. 

Our analysis has also shown that there is a quick technique for
deriving approximation methods to arbitrarily high order involving the
solution of relatively simple equations at each order. We have also
presented these results, but it turns out that they lead to
approximations that are far from optimal in the sense that there are
many more gates in these methods than should be necessary. That is,
they are accurate to high order, but relatively costly to implement.

As an example of how useful our approximations can be, let us consider
a case in which we want to apply an approximation method for time $T =
1$ with total error $E = 10^{-4}$. For a first order method, this
means that we require about $5000$ applications of the method. For
second order, we require about $30$ applications. For our third order
method ${\cal Z}_3^1$, we need $2$ applications. And for fourth order
method ${\cal Z}_4^1$, one application of the method is more than
sufficient. This results in a reduction of orders of magnitude in the
computational cost of a given simulation or gate application.

Using our equations, it is possible to search for 5th order methods
(and from these, via transposition, to obtain 6th order methods). We
made a number of attempts at the search, but were unable to find any
5th order methods due to the large size of the search space. Thus, the
only methods of 5th order and higher that we found were those methods
mentioned above which tend to involve unnecessarily large numbers of
gates.

\section*{Acknowledgements}

This work was supported by the DOE and NASA grant NAG 5-7092 at
Fermilab. We would like to thank Tasso Kaper for bringing references
\cite{otherwork} to our attention.

\newpage
\section*{Appendices}

\subsection*{A.1 \ \ Proof of lower bounds on integral method sizes}

\bea
\lt( \sigma_I^1 \rt)^p & = & \sum_{i=1}^I \lt[ \lt( \sigma_i^1 \rt)^p
- \lt( \sigma_{i-1}^1 \rt)^p \rt] \nn \\
& = & \sum_{i=1}^I \lt[ \lt( \sigma_{i-1}^1 + \alpha_i a_i \rt)^p -
\lt( \sigma_{i-1}^1 \rt)^p \rt] \nn \\
& = & \sum_{i=1}^I \lt[ \sum_{q=1}^{p-1} \frac{p!}{q! (p-q)!} \lt(\alpha_i
a_i \rt)^q \lt( \sigma_{i-1}^1 \rt)^{p-q} + \alpha_i^p a_i^p \rt]
\eea
$\alpha_i = \pm 1$. Therefore, if $p$ is odd, then
\be
  \sum_{i=1}^I \alpha_i^p a_i^p = \sum_{i=1}^I \alpha_i a_i^p = \sigma_I^p
\ee
and if $p$ is even, then
\be
  \sum_{i=1}^I \alpha_i^p a_i^p 
= \sum_{i=1}^I a_i^p 
= \sum_{i=1}^I \lt( 1-\alpha_i \rt) a_i^p + \sum_{i=1}^I \alpha_i a_i^p 
= \sum_{i=1}^I \lt( 1-\alpha_i \rt) a_i^p + \sigma_I^p
\ee

Taking $p = 2$, the factor $\frac{p!}{q! (p-q)!}$, $q = 1$, is equal
to 2, and the factor $\lt( 1-\alpha_i \rt)$ is 0 or 2. A 2nd-order
method requires $\sigma_I^2 = 0$, therefore $\lt(\sigma_I^1\rt)^2$
must be even, and so $\sigma_I^1$ must also be even.

Taking $p = 3$, the factors $\frac{p!}{q! (p-q)!}$, $q = 1,2$, are
equal to 3. A 3rd-order method requires $\sigma_I^3 = 0$,
therefore $\lt(\sigma_I^1\rt)^3$ must be a multiple of 3, and so
$\sigma_I^1$ must also be a multiple of 3.

Taking $p = 4$, the factor $\frac{p!}{q! (p-q)!}$, $q = 1, 2, 3$, is
even, and the factor $\lt( 1-\alpha_i \rt)$ is 0 or 2. A 4th-order
method requires $\sigma_I^4 = 0$, therefore $\lt(\sigma_I^1\rt)^4$
must be even, and so $\sigma_I^1$ must also be even.

Taking $p = 5$, the factors $\frac{p!}{q! (p-q)!}$, $q = 1, 2, 3, 4$,
are multiples of 5. A 5th-order method requires $\sigma_I^5 = 0$,
therefore $\lt(\sigma_I^1\rt)^5$ must be a multiple of 5, and so
$\sigma_I^1$ must also be a multiple of 5.

Combining these $\sigma_I^1$ must be a multiple of 2 in a 2nd-order
method, a multiple of 6 in a 3rd or 4th order method, and a multiple
of 30 in a 5th-order method.

\newpage
\subsection*{A.2 \ \ Formulae for the $B_2^X$'s in terms of the
commutators of $A_1$ and $A_2$}

The $B_2^p$ are defined by
\be
e^{a A_1} e^{a A_2} = \exp \sum_{p=1}^\infty a^p B_2^p
\ee
The Campbell-Baker-Hausdorff formula, Eq.~(\ref{CH}), then gives
\bea
B_2^1 &=& A_1 + A_2
\\
B_2^2 &=& \frac{1}{2} A_{12}
\\
B_2^3 &=& \frac{1}{12} \lt( A_{112} + A_{221} \rt)
\\
B_2^{12} \equiv \lt[ B_2^1 , B_2^2 \rt]
&=& \frac{1}{2} \lt( A_{112} - A_{221} \rt)
\\
B_2^4 &=& \frac{1}{24} A_{1221}
\\
B_2^{13} \equiv \lt[ B_2^1 , B_2^3 \rt]
&=& \frac{1}{12} \lt( A_{1112} + A_{2221} \rt)
\\
B_2^{112} \equiv \lt[ B_2^1 , B_2^{12} \rt]
&=& \frac{1}{2} \lt( A_{1112} - A_{2221} - 2 A_{1221} \rt)
\\
B_2^5 &=& - \frac{1}{720} \lt( A_{11112} - 2 A_{21112} - 6 A_{11221} \rt.
\nn \\
&& \lt. \hspace{2cm} \mbox{} - 6 A_{22112} - 2 A_{12221} + A_{22221} \rt)
\\
B_2^{14} \equiv \lt[ B_2^1 , B_2^4 \rt]
&=& \frac{1}{24} \lt( A_{11221} - A_{22112} \rt)
\\
B_2^{23} \equiv \lt[ B_2^2 , B_2^3 \rt]
&=& \frac{1}{24} \lt( \lt[ A_{12} , A_{112} \rt]
 + \lt[ A_{12} , A_{221} \rt] \rt) \nn
\\
&=& - \frac{1}{24} \lt( A_{21112} + A_{11221} - A_{22112} - A_{12221} \rt)
\\
B_2^{113} \equiv \lt[ B_2^1 , B_2^{13} \rt]
&=& \frac{1}{12} \lt( A_{11112} + A_{21112} + A_{12221} + A_{22221} \rt)
\\
B_2^{221} \equiv - \lt[ B_2^2 , B_2^{12} \rt]
&=& \frac{1}{4} \lt( - \lt[ A_{12} , A_{112} \rt]
 + \lt[ A_{12} , A_{221} \rt] \rt) \nn
\\
&=& \frac{1}{4} \lt( A_{21112} + A_{11221} + A_{22112} + A_{12221} \rt)
\\
B_2^{1112}
\equiv \lt[ B_2^1 , B_2^{112} \rt]
&=& \frac{1}{2} \lt( A_{11112} + A_{21112} - 2 A_{11221} \rt. \nn
\\
&& \lt. \hspace{2cm} \mbox{} + 2 A_{22112} - A_{12221} - A_{22221} \rt)
\eea

\newpage
\subsection*{A.3 \ \ A simple measure of the error}

The error for a given method is given by Eq.~(\ref{r})
\be
r = \sum_X \sigma_I^X B_N^X
\ee
where $X \in \{4, 13, 112\}$ for a 3rd order method and $X \in \{5,
14, 23, 113, 221, 1112\}$ for a 4th order method.

$r$ is a vector in the vector space of the commutators for which we do
not know the metric. We would like to have a scalar measure of the
error, and thus must pick some basis for the vector space. We choose
the basis to be the commutators of $A_1$ and $A_2$. This basis is
simple and spans the vector space of the $B_N^X$'s without
redundancy. Since this basis spans the space of the $B_N^X$'s, we
do not need to go to $N$ larger than $2$. For $N = 2$, we can
re-express $r$ as
\be
r = \sum_Y \rho_Y A_Y
\ee
where 
\be
Y \in \{1112, 1221, 2221\}
\ee
for a 3rd order method and
\be
Y \in \{11112, 21112, 11221, 22112, 12221, 22221\}
\ee
for a 4th order method. The formulae for the $\rho_Y$'s in terms of
the $\sigma_I^X$'s are given in Appendix A.4. In this basis, our
measure of the error then becomes
\be
R \equiv \sqrt{\sum_Y (\rho_Y)^2}
\ee

\newpage
\subsection*{A.4 \ \ Formulae for the $\rho_Y$'s in terms of the
$\sigma_I^X$'s}

For $N = 2$,
\be
  r = \sum_X \sigma_I^X B_2^X = \sum_Y \rho_Y A_Y
\ee
Therefore, using the formulae in Appendix A.2, we obtain

\bea
\rho_1 &=& \sigma_I^1
\\
\rho_2 &=& \sigma_I^1
\\
\rho_{12} &=& \frac{1}{2}\sigma_I^2
\\
\rho_{112} &=& \frac{1}{12}\sigma_I^3 + \frac{1}{2}\sigma_I^{12}
\\
\rho_{221} &=& \frac{1}{12}\sigma_I^3 - \frac{1}{2}\sigma_I^{12}
\\
\rho_{1112} &=& \frac{1}{12}\sigma_I^{13} + \frac{1}{2}\sigma_I^{112}
\\
\rho_{1221} &=& \frac{1}{24}\sigma_I^{4} - \sigma_I^{112}
\\
\rho_{2221} &=& \frac{1}{12}\sigma_I^{13} - \frac{1}{2}\sigma_I^{112}
\\
\rho_{11112} &=& -\frac{1}{720}\sigma_I^{5} 
          + \frac{1}{12}\sigma_I^{113} + \frac{1}{2}\sigma_I^{1112} 
\\
\rho_{21112} &=& \frac{1}{360}\sigma_I^{5} - \frac{1}{24}\sigma_I^{23} 
          + \frac{1}{12}\sigma_I^{113} + \frac{1}{4}\sigma_I^{221} 
          + \frac{1}{2}\sigma_I^{1112} 
\\
\rho_{11221} &=& \frac{1}{120}\sigma_I^{5} + \frac{1}{24}\sigma_I^{14} 
          - \frac{1}{24}\sigma_I^{23} + \frac{1}{4}\sigma_I^{221} 
          - \sigma_I^{1112} 
\\
\rho_{22112} &=& \frac{1}{120}\sigma_I^{5} - \frac{1}{24}\sigma_I^{14} 
          + \frac{1}{24}\sigma_I^{23} + \frac{1}{4}\sigma_I^{221} 
          + \sigma_I^{1112} 
\\
\rho_{12221} &=& \frac{1}{360}\sigma_I^{5} + \frac{1}{24}\sigma_I^{23} 
          + \frac{1}{12}\sigma_I^{113} + \frac{1}{4}\sigma_I^{221} 
          - \frac{1}{2}\sigma_I^{1112} 
\\
\rho_{22221} &=& -\frac{1}{720}\sigma_I^{5} + \frac{1}{12}\sigma_I^{113}
          - \frac{1}{2}\sigma_I^{1112} 
\eea

\newpage
\subsection*{A.5 \ \ Tables of Residual Errors}

3rd Order Integer Methods

\begin{flushleft}
\begin{tabular}{|r|r|r|r|r|}
\hline
& $\rho_{1}$ 
& $\rho_{1112}$ & $\rho_{1221}$ & $\rho_{2221}$ \\
\hline
${\cal Z}_3^1$ & 6.0
  & -1.0 & 0.5 & 0.0 \\
\hline
${\cal Z}_3^2$ & 12.0
  & -4.0 & -3.0 & 5.0 \\
\hline
${\cal Z}_3^3$ & 6.0
  & -2.0 & 1.5 & 1.0 \\
\hline
${\cal Z}_3^4$ & 6.0
  & 0.0 & 4.5 & 9.0 \\
\hline
${\cal Z}_3^5$ & 12.0
  & -864.0 & 792.0 & 180.0 \\
\hline
\end{tabular}

\vspace{2ex}

\begin{tabular}{|r|r|r|r|r|r|r|}
\hline
& $\rho_{11112}$ & $\rho_{21112}$ & $\rho_{11221}$ 
& $\rho_{22112}$ & $\rho_{12221}$ & $\rho_{22221}$ \\
\hline
${\cal Z}_3^1$ & 2.2 & 3.1 & -3.2
  & 5.3 & 0.1 & -1.3 \\
\hline
${\cal Z}_3^2$ & 13.4 & 104.2 & 105.6
  & 26.1 & 84.2  & 28.9 \\
\hline
${\cal Z}_3^3$ & 0.7 & 5.1 & 3.3
  & 1.8 & 3.1 & 1.2 \\
\hline
${\cal Z}_3^4$ & 2.7 & 8.1 & -2.7
  & 10.8 & -6.9 & -13.8 \\
\hline
${\cal Z}_3^5$ & -3801.6 & -1900.8 & 2505.6
  & -1166.4 & 499.2 & 206.4 \\
\hline
\end{tabular}

\vspace{3ex}

3rd Order Irrational Method

\vspace{2ex}

\begin{tabular}{|r|r|r|r|r|}
\hline
& $\rho_{1}$ 
& $\rho_{1112}$ & $\rho_{1221}$ & $\rho_{2221}$ \\
\hline
${\cal R}_3^1$ & 1.0
  & 0.012008 & -0.052816 & -0.058414 \\
\hline
\end{tabular}

\vspace{2ex}

\begin{tabular}{|r|r|r|r|r|r|r|}
\hline
& $\rho_{11112}$ & $\rho_{21112}$ & $\rho_{11221}$ 
& $\rho_{22112}$ & $\rho_{12221}$ & $\rho_{22221}$ \\
\hline
${\cal R}_3^1$ & 0.001754 & 0.003500 & -0.009304
  & 0.017412 & -0.014311 & -0.026310 \\
\hline
\end{tabular}

\vspace{3ex}

4th Order Integer Methods

\vspace{2ex}

\begin{tabular}{|r|r|r|r|r|r|r|r|}
\hline
& $\rho_1$ 
& $\rho_{11112}$ & $\rho_{21112}$ & $\rho_{11221}$ 
& $\rho_{22112}$ & $\rho_{12221}$ & $\rho_{22221}$ \\
\hline
${\cal R}_4^1$ & 12.0
  & -1.6 & 0.2 & -3.4
  & 5.6 & -1.8 & -2.6 \\
\hline
${\cal R}_4^2$ & 12.0
  & 3.4 & 6.2 & 3.6
  & 3.6 & 2.2 & -4.6 \\
\hline
${\cal R}_4^3$ & 12.0
  & 26.4 & 40.2 & -5.4
  & 21.6 & 16.2 & 5.4 \\
\hline
${\cal R}_4^4$ & 12.0
  & -369.6 & -220.8 & 309.6
  & -86.4 & 259.2 & 86.4 \\
\hline
\end{tabular}

\vspace{3ex}

4th Order Irrational Methods

\vspace{2ex}

\begin{tabular}{|r|r|r|r|r|r|r|r|}
\hline
& $\rho_1$ 
& $\rho_{11112}$ & $\rho_{21112}$ & $\rho_{11221}$ 
& $\rho_{22112}$ & $\rho_{12221}$ & $\rho_{22221}$ \\
\hline
${\cal R}_4^1$ & 1.0
  & -0.000414 & -0.008682 & -0.007027
  & -0.026045 & -0.026732 & -0.004684 \\
\hline
${\cal R}_4^2$ & 1.0
  & -0.022171 & -0.013256 & 0.014902
  & -0.009176 &  0.002796 & 0.001717 \\
\hline
${\cal R}_4^3$ & 1.0
  & -0.001297 &  0.038072 & 0.035227
  & -0.080082 & -0.079215 & 0.001270 \\
\hline
${\cal R}_4^4$ & 1.0
  &  0.002074 &  0.196582 &  0.194095
  & -0.052861 & -0.050727 & -0.002155 \\
\hline
\end{tabular}
\end{flushleft}


\begin{thebibliography}{99}

\bibitem{shor} P. W. Shor, in {\it Proceedings of the 35th Annual
Symposium on Foundations of Computer Science, Santa Fe, NM, 1994},
edited by Shafi Goldwasser (IEEE Computer Society Press, Los Alamitos,
CA, 1994), 124-134; SIAM J. Comput. {\bf 26}, 1484-1509 (1997).

\bibitem{preskill} J. Preskill, quant-ph/9712048.

\bibitem{zurek} R. Laflamme, E. Knill, W. H. Zurek, T. F. Havel and
S. S. Somaroo, {\it Phys. Rev. Lett.} {\bf 81}, 2152-2155 (1998).

\bibitem{grover} L. Grover, {\it Phys. Rev. Lett.} {\bf 79}, 325-328
(1997).

\bibitem{simulate} B. M. Boghosian and W. Taylor IV, quant-ph/9701019;
quant-ph/9701016; quant-ph/9604035; D. S. Abrams and S. Lloyd, {\it
Phys. Rev. Lett.} {\bf 79}, 2589-2589 (1997). 

\bibitem{feynman} R. P. Feynman, {\it Int. Jour. of Theor. Phys.}
{\bf 21}, 467-488 (1982).

\bibitem{abramslloyd}  D. S. Abrams and S. Lloyd, {\it
Phys. Rev. Lett.} {\bf 79}, 2586-2589 (1997).

\bibitem{qc} A. Steane, quant-ph/9708022; S. Lloyd, {\it
Phys. Rev. Lett.} {\bf 75}, 346-349 (1995); D. Deutsch, A. Barenco and
A. Ekert, quant-ph/9505018; A. Barenco, et. al., quant-ph/9503016.

\bibitem{otherwork} D. Goldman and T. Kaper, {\it SIAM J. Num. Anal.}
{\bf 33}, 349-367 (1996). M. Suzuki, {\it Phys. Lett. A}
{\bf 146}, 319-323 (1990); M. Suzuki, {\it J. Math. Phys.}
{\bf 32}, 400-407 (1991);

\bibitem{numrec} {\it Numerical Recipes, The Art of Scientific
Computing}, W. H. Press, S. A. Teukolsky, W. T. Vetterling and
B. P. Flannery, Cambridge University Press, 376-381 (1992).

\bibitem{yoshida} H. Yoshida, {\it Phys. Lett. A} {\bf 150}, 262-268
(1990).

\end{thebibliography}
\end{document}